\documentclass{PoS}

\usepackage{amsmath}
\usepackage{amsfonts}
\usepackage{amssymb}
\usepackage{graphicx}
\usepackage{multirow}
\usepackage{multicol}
\usepackage{tabularx}
\usepackage{booktabs}
\usepackage{color}
\usepackage{etoolbox}
\usepackage{lipsum}
\usepackage{xcolor}
\usepackage{changepage}
\usepackage{gensymb}
\usepackage[normalem]{ulem}
\usepackage{comment}
\usepackage{mathrsfs}

\let\OLDthebibliography\thebibliography
\renewcommand\thebibliography[1]{
  \OLDthebibliography{#1}
  \setlength{\parskip}{0pt}
  \setlength{\itemsep}{0pt plus 0.3ex}
}

\title{VHE gamma-ray study of the composite \\SNR MSH 15-5\textit{2} with H.E.S.S.}

\ShortTitle{Study of the VHE emission of the PWN in MSH 15-5\textit{2}}

\author{\speaker{M. Tsirou} \\%
        LUPM, CNRS-IN2P3 / University of Montpellier, France\\
        E-mail: \email{michelle.tsirou@etu.umontpellier.fr}}

\author{Y. A. Gallant\\
        LUPM, CNRS-IN2P3 / University of Montpellier, France\\
        E-mail: \email{gallant@in2p3.fr}}
        
\author{R. Terrier\\
         APC, CNRS-IN2P3 / Paris VII University, France\\
        E-mail: \email{rterrier@apc.in2p3.fr}}

\author{R. Zanin\\
        MPI-K / Heidelberg, Germany\\
        E-mail: \email{Roberta.Zanin@mpi-hd.mpg.de}}
        
\author{for the H.E.S.S. Collaboration\\}

\abstract{The composite supernova remnant (SNR) MSH 15-5\textit{2} comprises the bright X-ray pulsar wind nebula (PWN) of PSR B1509-58, surrounded by a shell which is a prominent object in the radio domain. H.E.S.S. had discovered extended very-high-energy (VHE) gamma-ray emission coincident with the PWN. With additional H.E.S.S. observations performed since the 2005 discovery paper, we study the properties of the emission in greater detail. We compare the VHE gamma-ray morphology of the PWN with that in synchrotron emission, obtained from archival X-ray observations, and discuss the implications on the magnetic field in the nebula. In particular, we discuss potential extended gamma-ray emission beyond the X-ray PWN, which may allow for conclusions on scenarios of PWNe as sources of cosmic ray electrons and positrons.}

\FullConference{35th International Cosmic Ray Conference $–$ ICRC2017-\\
		10-20 July, 2017\\
		Bexco, Busan, Korea}

\begin{document}

\section{Introduction}

\setcounter{page}{2} 

Pulsar Wind Nebulae (PWNe) are magnetized clouds of relativistic particles thought to be accelerated by a pulsar, that may be found in the inner regions of a supernova remnant (SNR). The composite SNR MSH 15-5\textit{2} and the central compact object that has been identified as the pulsar PSR B1509-58 have been observed in X-rays and also in radio wavelengths. In 2005, VHE $\gamma$-ray emission was discovered by the High Energy Stereoscopic System (H.E.S.S.), using 22.1h of data, and the emission morphology was described as an elliptical Gaussian$^{\cite{2005HESS}}$.

In this proceeding, we use a larger dataset of H.E.S.S. observations, from 2004 until 2014, consisting of the H.E.S.S. phase I era  where events are reconstructed from an array of four 12-m sized Cherenkov telescopes. We have more than twice the statistics of the discovery paper, in order to disentangle the morphology of MSH 15-5\textit{2}. Furthermore, we use X-ray \textit{Chandra} observations so as to compare the VHE emission produced via Inverse Compton in $\gamma$-rays with the corresponding synchrotron emission of the PWN. We show that in order to model well the emission, one needs to modify the model using the X-ray template, linked to the assumptions one makes on the magnetic field. Moreover we show that an additional geometrical component is required in order to describe the extended $\gamma$-ray emission beyond the outskirts of the X-ray PWN. Finally we hypothesize, given its shape, on the behaviour of the leptonic population within this object. 

\section{Morphological analysis}

\subsection{Gamma-ray emission}
\label{configuration}
In order to analyse the morphology of the composite MSH 15-5\textit{2}, we use $\gamma$-ray count maps extracted with the \textit{ParisAnalysis} analysis chain\footnote{\samepage using the HESS-Soft-0-8-24 version, Prod26 DSTs} using a Model ++ standard reconstruction of events. The radius of selection on the given runs was set at 2.5$\degree$ and standard cuts were chosen ($\geq$ 60 p.e., corresponding to $\gtrsim$ 0.3 TeV) resulting in H.E.S.S. I data with a live time of 95.9 hours (48.5~h exposure-corrected). The results were cross-checked by performing a similar image analysis on maps obtained by an independent calibration and reconstruction chain.

The fitting procedure was made with the Sherpa package$^{\cite{Sherpa}}$, using the raw count map of gamma-like events, the Point Spread Function image (PSF), a smoothed background and an integrated exposure map assuming a spectral index $\Gamma$ = 2.27$^{\cite{2005HESS}}$.  

\subsection{X-ray emission}
\label{xrays}
\subsubsection{Emission from leptons in PWNe}
\label{energies}
In PWNe, the more common mechanism of emission up to the TeV range would be Inverse Compton (IC) scattering of seed photons by leptons. The seed photon population may be mainly composed by the Cosmic Microwave Background (CMB), the Galactic Far Infra-Red background (FIR) and the Galactic stellar background may also contribute. As we are interested in the lepton distribution in the source, we examine to what extent the leptons responsible for $\gamma$-ray IC emission correspond in energy range to those which produce synchrotron radiation in the X-ray range.
One-zone model fits of the synchrotron and IC emission suggest a magnetic field value $B \approx17\,\mu$G$^{\cite{2005HESS,Fermi10}}$ in the PWN; electrons whose synchrotron spectrum peaks at $h\nu_{\rm synch} > 4$\,keV then have energies $E_{e} \gtrsim 100$\,TeV. The same models of the $\gamma$-ray IC emission suggest that the Galactic FIR background is the dominant target photon component$^{\cite{2005HESS,Fermi10}}$. Thus for a blackbody temperature of $\sim$ 45 K, $\gamma$-rays with $E_{\gamma} \gtrsim$ 0.3 TeV are predominantly scattered by leptons with $E_{e} \gtrsim 2$\,TeV.
While the $\gamma$-ray morphology thus traces leptons of slightly lower energy than the X-ray synchrotron map, it is plausible to assume that they follow the same distribution given the compatible spectral indices$^{\cite{Gaeetc02}}$.

\begin{figure}[t]
	\begin{adjustwidth}{+6em}{+6em}

			\textit{Radio band (843 MHz)}
			\hspace{0.8cm}
			\textit{Synchrotron X-ray template (4 - 7 keV)}
	
		\includegraphics[width=0.4\linewidth, height=45mm]{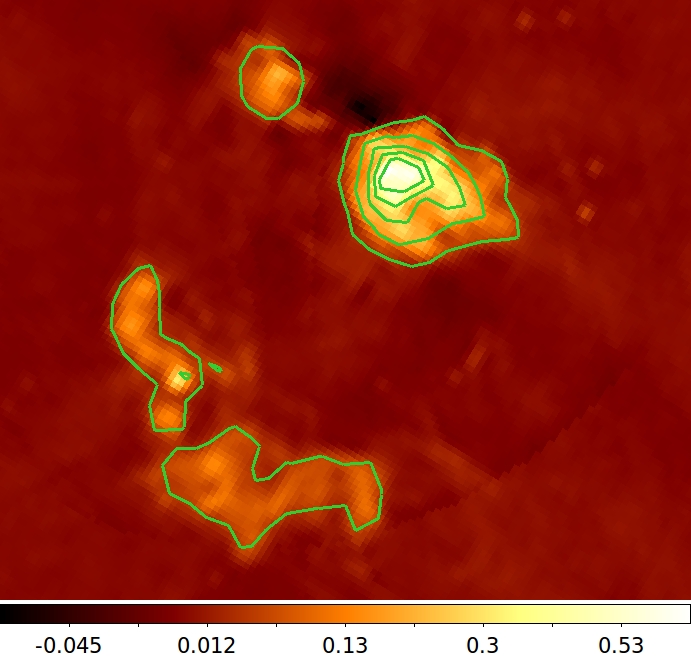}
		\includegraphics[width=0.4\linewidth, height=45mm]{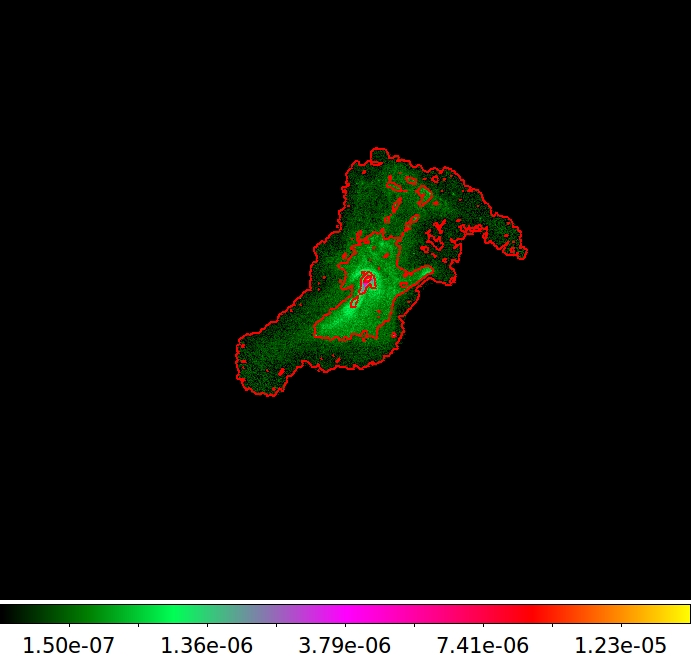}
		\\~\\
			\textit{$\gamma$-ray band ($\geq$ 0.3 TeV)}
			
		\includegraphics[width=0.4\linewidth, height=45mm]{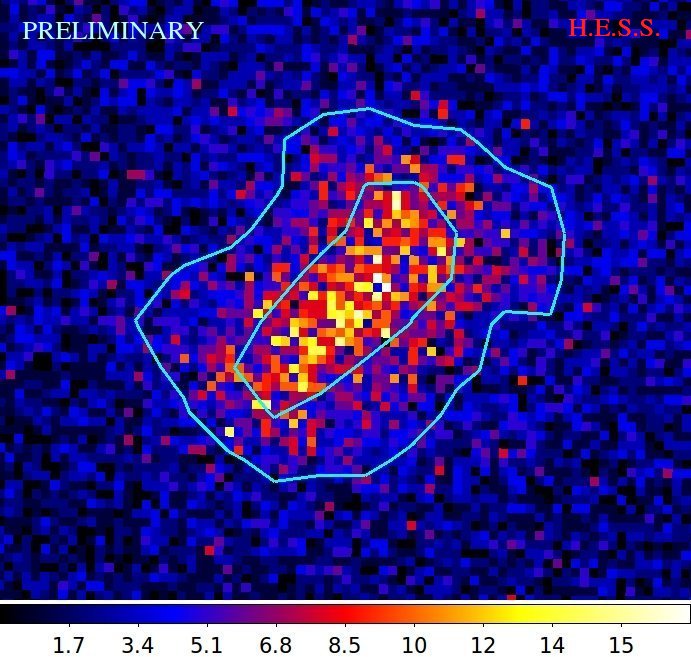}
		\includegraphics[width=0.45\linewidth, height=45mm]{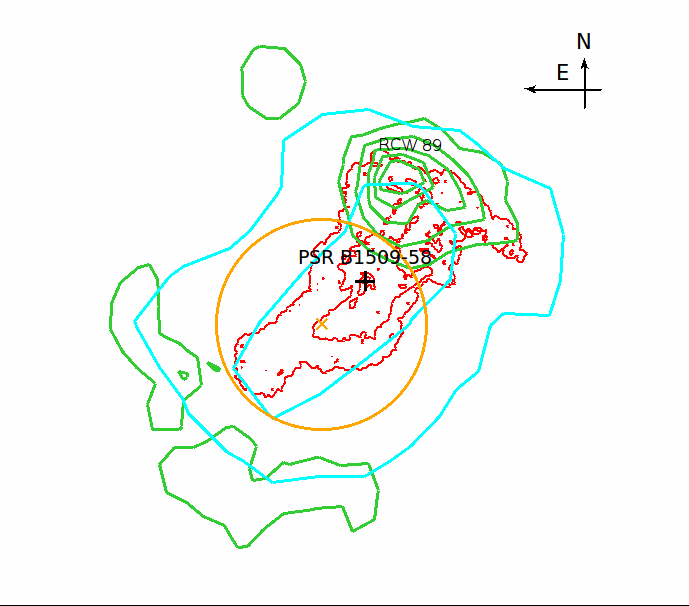}
\begin{flushright}
			\textit{Contours in the radio, X-ray and $\gamma$-ray bands}

\end{flushright}
	\end{adjustwidth}
	\caption{Maps in different wavelengths for the composite SNR MSH 15-5\textit{2}. \textit{Top left} : radio band. Image extracted from the Molonglo Galactic Plane Survey (MGPS) using the Molonglo Observatory Synthesis Telescope (MOST) at 843 MHz with a resolution of $43'' \times 43''$ cosec $|$dec$|$. The green contours outline parts of the shell of the SNR MSH 15-5\textit{2}. \textit{Top right} : X-ray template extracted from four Chandra archival observations ($\sim$ 200 ks) in the 4 to 7 keV band. The red contours are shown as a tracer of the PWN size. \textit{Bottom left} : $\gamma$-ray emission from $\sim$0.3 TeV and above detected by H.E.S.S.-I. The cyan contours outline the peaked central VHE emission and the more extended one. \textit{Bottom right} : The black cross indicates the position of the pulsar PSR B1509-58. The cyan contours are drawn for the $\gamma$-ray emission, the red contours for the X-ray emission and the green ones for the radio emission. The orange circle represents the extent of the symmetrical added Gaussian component found in this analysis for the model \textit{v1R+G}. (see Sect.~\ref{v1r+})}
\label{multiwavelengths}

\end{figure}

\subsubsection{X-ray template}

For the morphological fitting of MSH 15-5\textit{2}, we use four archival X-ray \textit{Chandra} observations resulting in a total of $\sim$\,200\,ks.  We choose the energy range 4--7\,keV, in order to omit as much as possible the thermal emission at lower energy from the SNR, and in particular from the H$\alpha$ nebula RCW 89 in the NW part of the image (see Fig.\ref{multiwavelengths}). Spectroscopic imaging suggests that the thermally-emitting clumps in RCW 89, which may be due to the encounter of the pulsar jet with the SNR shell boundary, show no detectable emission
above $\sim$\,4\,keV$^{\cite{Gaeetc02}}$.

As a template for the synchrotron map in the morphological fit, we used a modified version of the exposure-corrected X-ray map. An average estimate of the background, based on four regions in the image, was subtracted from the template, and the emission coming from the pulsar has been replaced by an average of its immediate surroundings. Furthermore, a mask has been applied in order to keep only the significant X-ray emission from the PWN and to prevent noise at the outskirts of the image from influencing the fit.

\section{Results}
\subsection{Fitting models}
\label{models}

For the fitting procedure, the different models consisted of first fixing the background, then convolving the X-ray template (and the additional components) with the PSF and multiplying the result by the exposure map. The influence of the PSF has been explored by testing our procedure using a PSF broadened with a standard deviation of $\sigma$ = 0.022$\degree$, which is representative of the PSF uncertainty based on a systematic study.

\subsubsection{Model v1 : X-ray template}
\label{v1}
When using only the X-ray template for fitting the shape of the gamma candidates count map, the "\textit{v1}" model, the results indicate that the central region around the pulsar is dominating the fit procedure, even though the emission coming from the pulsar itself had been previously removed in the image analysis. This can be clearly seen in the residual map, where there is an apparent correlation between the position of the pulsar and the striking negative residuals (see Fig.~\ref{image_results}, left column). Thus the actual longitudinal profile of MSH 15-5\textit{2} is not fully reproduced, as in addition to the plunging negative residuals mentioned above, there are also positive residuals at larger distances from the central emission. 

\subsubsection{Model v1R : X-ray template $\times$ R$^{\alpha}$}
\label{v1R}
The previous model (Sect.~\ref{v1}) has shown that the X-ray emission coming from the region around the pulsar is still too bright when compared to our $\gamma$-ray data. Therefore, by assuming that the magnetic field \textbf{B} varies spatially, we imposed a dependency of \textbf{B} with the projected distance $R$ to the pulsar position, by multiplying the X-ray template by $R^{\alpha}$ where $\alpha$ is a given index to be determined by the fit results. Let us call this model "\textit{v1R}". This parameter $\alpha$, should hint at the actual dependency of $\textbf{B}$ with the distance $R$. Thus based on this magnetic field component approximation, the Inverse Compton flux $F_{IC}$ is approximated to the synchrotron flux $F_{sync}$ times the distance $R$ to a given power $\alpha$ : $F_{IC}$ $\propto$ $F_{synch}$ $\times$ $R^{\alpha}$. As for the mentioned fluxes, $F_{sync}$ is proportional to the lepton density $N_{e}$ and to the magnetic field $B$ with a given index $\beta$ : $F_{synch}$($\nu$) $\propto$ $B^{\beta}$ $\times$ $N_{e}$. In turn, B is dependent to the distance with another given index $\delta$ : $B \propto$ $R^{\delta}$. On the other hand, $F_{IC}$ is proportional to $N_{e}$ and the photon density $U_{\phi}$ which is assumed to be uniform. With these different approximations in hand, one may easily find that $\frac{F_{IC}}{ F_{synch}} \propto$ $R^{-\beta\delta}$ $\propto$ $R^{\alpha}$ thus giving the following simple relation between those indices : $\delta$ = - $\frac{\alpha}{\beta}$ .

The distribution of the leptonic population can be described as a power law such as $N(E) $d$E \propto \kappa\, E^{-p} $d$E$. The dependency of the flux with the frequency $\nu$ and the magnetic field is then such as $F_{\nu} \propto \nu^{\frac{1}{2} (1 - p)} B^{\frac{1}{2} (1 + p)}$. Thus we can link the previously mentioned index  $\beta$ with $p$ the electron index and $s$ the synchrotron index such as $\beta = \frac{1}{2} (1 + p) = s + 1$. Thanks to this relation, when using our fitted estimations of the $\alpha$ index $\in$ [1.0 , 1.3] then for a spectral index of $\beta \approx $ 2.2, one finds that the dependency between the magnetic field and the distance is $B \propto R^{\delta}$ with $\delta$ $\in$ [- 0.4 , -0.6].

\subsubsection{Model v1R+G : X-ray template $\times$ R$^{\alpha}$ + Gaussian component}
\label{v1r+}
\begin{figure}[t]
	\begin{adjustwidth}{-1.6em}{-1.6em}

		\includegraphics[width=0.33\linewidth, height=43mm]{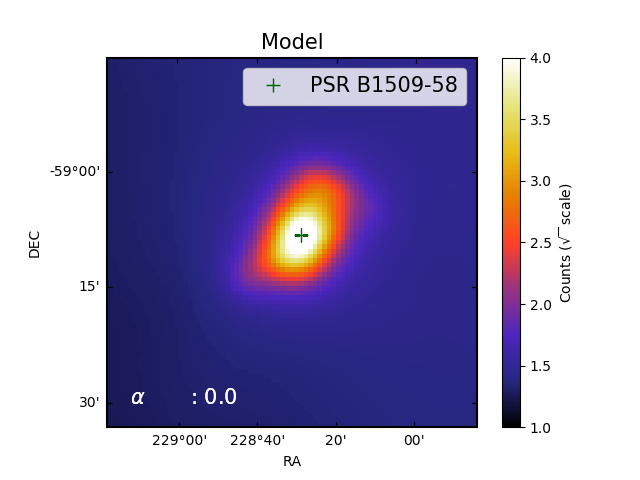}
		\includegraphics[width=0.33\linewidth, height=43mm]{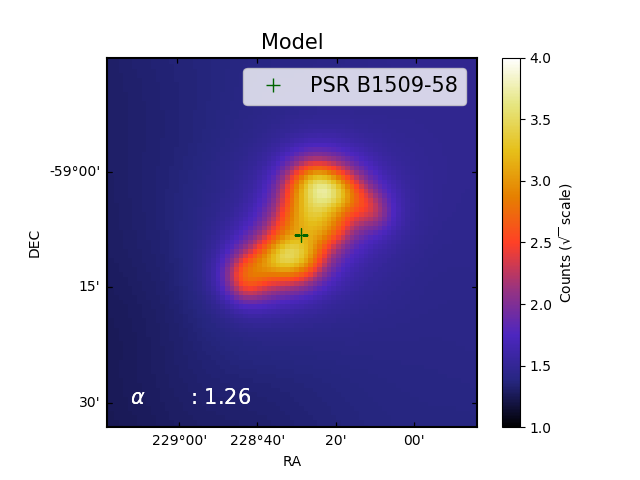}
		\includegraphics[width=0.33\linewidth, height=43mm]{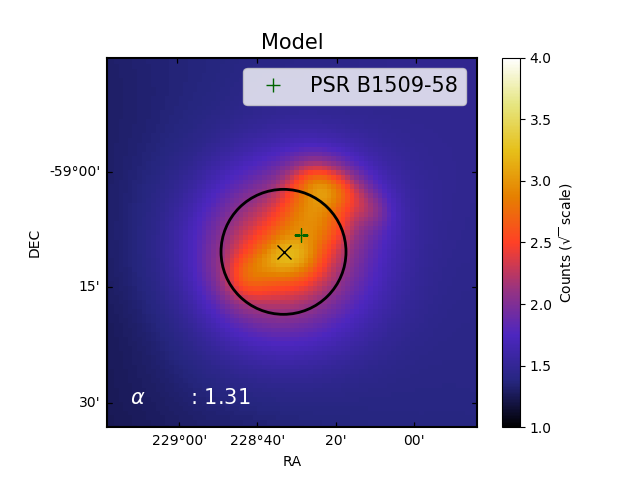}

		\includegraphics[width=0.33\linewidth, height=43mm]{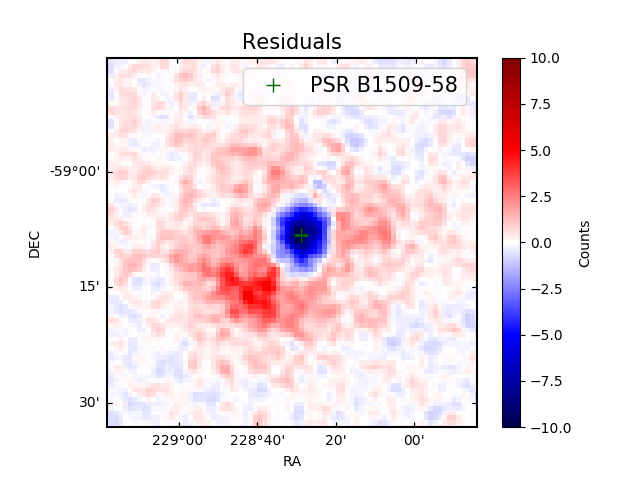}
		\includegraphics[width=0.33\linewidth, height=43mm]{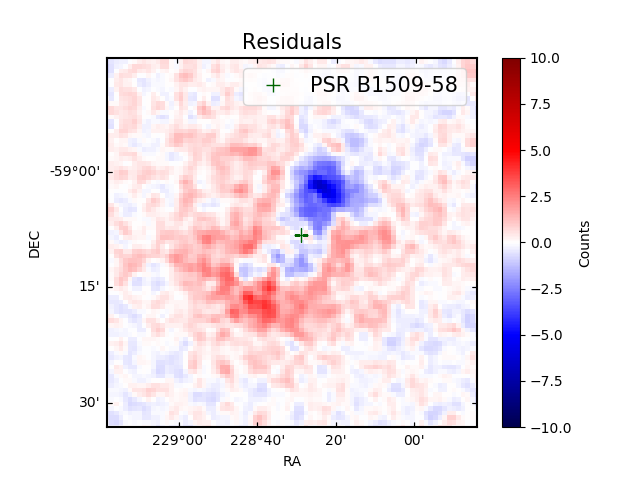}
		\includegraphics[width=0.33\linewidth, height=43mm]{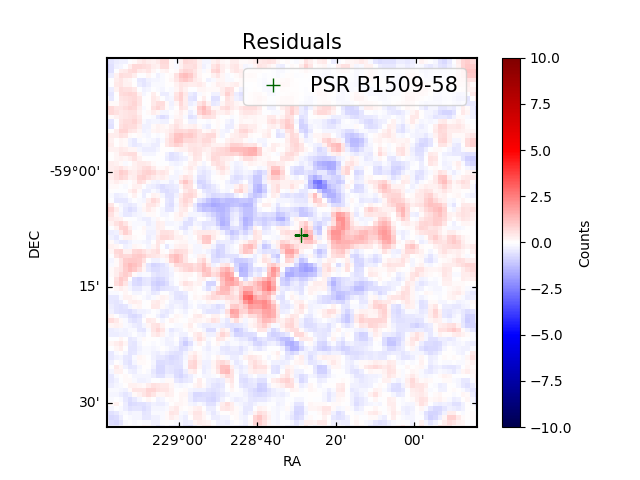}

				\hspace{0.7cm}
				\textit{v1 : X-ray template}
				\hspace{1.8cm}
				\textit{v1R : X-ray template $\times$ R$^{\alpha}$}
				\hspace{0.4cm}				
				\textit{v1R+G : X-ray template $\times$ R$^{\alpha}$ + G$_{sym}$}	
	
	\end{adjustwidth}
	\caption{Preliminary maps of the models (top) and the residuals (bottom) for the different morphological fits of the composite SNR MSH 15-5\textit{2}. From left to right : \textit{v1} model, \textit{v1R} model, \textit{v1R+G} model (see Sect.~\ref{models}). On the images, the label consists of the $\alpha$ parameter. The green cross represents the position of the pulsar PSR B1509-58. The black circle for the \textit{v1R+G} model represents the one-$\sigma$ extent of the symmetrical Gaussian component (see Sect.~\ref{v1r+}).}
	\label{image_results}
\end{figure}

As seen in Fig.~\ref{image_results} (middle column), the positive residuals between the \textit{v1R} model and the dataset indicate strongly that one needs to take into account another component in the model, hinting that it should be located around the southern part of the PWN in MSH 15-5\textit{2}. Therefore, we added a geometrical component to our previous model : a symmetrical Gaussian component. We also explored the possibility to add other types of components such as 
a shell or a disk. The results of the fitting procedures indicate that the Gaussian component better suits our data. This model will be mentioned as the "\textit{v1R+G}" model for the rest of the discussion.

\subsection{Statistical comparison}

In order to efficiently compare all of our models (nested or not), we use the Akaike Information Criterion (AIC)$^{\cite{Akaike}}$. It is a statistical tool comprising a likelihood function, also considering the number $k$ of independently adjusted parameters in the given model thus enabling the comparison between models with different number of parameters : AIC $= -2\log(L) + 2k$. Here the difference between the criteria $\Delta$AIC = AIC$_{\rm model\,1}$ - AIC$_{\rm model\,2}$ is used for determining the best model.

Our results (Table \ref{table_results}) indicate, based on the statistical test value, the residual maps and the stability of the convergence of the fit procedure, that the favored model for adequately describing the distribution of the VHE gamma ray emission coming from the composite SNR MSH 15-5\textit{2} is the one using the X-ray template alongside with the $\alpha$ parameter illustrating the magnetic field - distance dependency of the synchrotron emission and with an added geometrical component, being a symmetrical Gaussian (model "\textit{v1R+G}"). The intrinsic radius of this Gaussian component, which comprises about 65$\%$ of the total flux, is $\sim$\,$7 '$, implying that the $\gamma$-ray emission extends farther than the PWN observed in X-rays. In the following section we discuss possible interpretations of this component by introducing some scenarios concerning the behavior of the leptons. 

\begin{table}
\small
\begin{adjustwidth}{-1.8em}{1.8em}
\begin{tabularx}{1.05\linewidth}{l|ccc}
\toprule
\toprule
Model & v1 :  & v1R :  & v1R+G :  \\
      & X-ray template & X-ray template $\times$ R$^{\alpha}$ & X-ray template $\times$ R$^{\alpha}$ + component\\
\hline

$\alpha$ parameter &  0.00  &   1.26 $\pm$ 0.06$_{\rm stat}$ &  1.31 $\pm$ 0.14$_{\rm stat}$\\
\hline
AIC & 255 074 & 255 474 & 256 466 \\
\hline
$\Delta$AIC & 0 & 400 & 1 392 \\
\hline
X-ray amplitude (a.u.) & (1.5 $\pm$ 0.03$_{\rm stat}$) 10$^{5}$ & (1.7 $\pm$ 0.03$_{\rm stat}$) 10$^{5}$& (0.8 $\pm$ 0.05$_{\rm stat}$) 10$^{5}$ \\
\hline
\hline
Added component & no & no & yes : symmetrical Gaussian\\
\hline
Portion of the total flux  & & & 64.8 $\%$ $\pm$ 5.3$_{\rm stat}\%$ \\
\hline
RA & & & 15 h  14 m  18.0 s $\pm$ 2.1$_{\rm stat}$ s $\pm$ 2.6$_{\rm syst}$ s \\

Dec & & & $-59\,\degree$  $10\,'$  $59\,''$ $\pm$ $16\,_{\rm stat}'' \pm 20\,_{\rm syst}''$ \\
\hline
Extent & & & $\sigma$ = $6.89\,' \pm 0.20\,_{\rm stat}' \pm 0.45\,_{\rm syst}'$ \\

\hline
\bottomrule
\end{tabularx}
\caption{Preliminary results for the different fit models of the analysis with standard cuts after selecting the runs within 2.5$\degree$. The $\Delta$AIC is computed by comparison to the \textit{v1} model (only using the X-ray template). The portion of the total flux is computed as the fraction of the Gaussian integral over the total model flux. }
\label{table_results}
\end{adjustwidth}
\end{table}

\section{Discussion on the models}
\subsection{Composite SNR size and age}

   The characteristic spin-down time of PSR\,B1509--58 is $\tau \approx
1600$\,yr$^{\cite{psrcat}}$, which sets an upper limit to the pulsar age,
unless it had an unconventional spin-down history.
From H{\sc i} absorption measurements, Gaensler et al.$^{\cite{Gaeetc99}}$
conclude that the distance to the SNR lies in the range $3.8\,{\rm kpc} < D
< 6.6\,{\rm kpc}$, while the dispersion measure of the pulsar suggests a distance
$D \approx 4.4$\,kpc with the most recent Galactic free electron density
model$^{\cite{psrcat}}$.  In the following, we will normalize $D$ to a
fiducial distance of 4\,kpc: $d_4 \equiv D/(4\,{\rm kpc})$.

   The lack of measured pulsar proper motion suggests that the explosion
center was near the current position of PSR\,B1509--58. The SNR blast
wave must then have expanded asymmetrically, with a large radius of
$R = (20 \pm 2)\,d_4\,$pc being reached to the SE$^{\cite{Gaeetc99}}$.
To reconcile this large size with the relatively young age implied by
the $\tau$ of the pulsar, one must assume a fairly energetic explosion and
a low medium density on the SE side. Gaensler et al.$^{\cite{Gaeetc99}}$ suggest that the progenitor of this SNR was a helium
star, combining an ejecta mass $M_{\rm ej} \approx 1.5 M_\odot$ and an
explosion energy $E \sim 3 \times 10^{51}$\,erg, and that the circumstellar
medium density to the SE is $n_{\rm H} \sim 0.01\,{\rm cm}^{-3}$.

\subsection{Reverse shock interaction and offset relic PWN}
\label{shock}
   The assumed low density to the SE would imply that the blast wave
is still in the early part of the transition from the free expansion
to the Sedov-Taylor phase, and thus that the reverse shock on that
side has only traveled a small fraction of the distance back to the
center. 
On the NW side of the remnant, by contrast, the
much smaller forward shock radius can be explained by interaction
with a much denser medium, for which a density $n_{\rm H} \sim 1$--$5\,
{\rm cm}^{-3}$ can be estimated$^{\cite{Gaeetc99}}$. On the NW side, the
SNR would then be well into the Sedov-Taylor phase, with the reverse
shock having reached the center and interacted with the PWN. Blondin
et al.$^{\cite{BloCheFri01}}$ have shown that the evolution of a composite
SNR in such an asymmetric medium results in the reverse shock compressing
and displacing the PWN in the direction opposite the high density region,
yielding an offset, "relic" PWN comprising the particles injected by
the pulsar prior to the reverse shock interaction. This offset has been also observed in other "middle-aged" PWNe$^{\cite{pwn}}$ thus raising the
possibility that the additional Gaussian component in $\gamma$-rays which
we find to the SE of the pulsar might originate from such a relic PWN.

\subsection{Spectral break and cutoff}
\label{spectra}
As discussed in Sect.~\ref{energies}, the leptons responsible for the
$\gamma$-ray emission are of somewhat lower energy than those traced
by our synchrotron X-ray template.
With an assumed magnetic field value of 20 $\mu$G, the latter have
$E_e \gtrsim 100$\,TeV and radiative synchrotron lifetimes
$\lesssim 300$\,yr, considerably less than the age of the SNR,
and thus the X-ray morphology reflects recently-injected electrons.
The lack of a prominent extended Gaussian component in X-rays could
then be due to the relic nebula having a synchrotron cooling cutoff
at lower energies.  Such a steeper spectrum for the Gaussian component
could be confirmed by energy-dependent morphology in the H.E.S.S.
energy range; we note that the morphology observed with {\it Fermi}-LAT$^{\cite{Fermi10}}$ is compatible with our fitted Gaussian component.
An alternate possibility is that the electrons responsible for the
Gaussian component have a similar spectrum to the rest, but do not
radiate significant synchrotron X-rays because they are in a lower
magnetic field medium, such as the SN ejecta.

\subsection{Leptons escaping into ejecta}
\label{escape}
	The morphology of the PWN in the X-ray band (Fig.~\ref{multiwavelengths}) appears somewhat irregular. The nebula follows the rotation axis of the pulsar, however there is a strong asymmetry between the NW and SE regions, alongside the jets of the pulsar. If the halo and the other component share the same spectral behaviour, then one could have an estimation of the magnetic field. The fact that the synchrotron radiation does not match the extent of the $\gamma$-ray emission points to a relatively low magnetic field medium surrounding the PWN, such as the SN ejecta. In order to fully study this hypothesis, we would need to take into account the somewhat complex geometry of the system resulting by the reverse shock interacting with the ejecta. This also implies a strongly diffusive medium, with an energy-dependent diffusion coefficent $\kappa$ at $\sim$ 10$^{28}$ cm$^{2}$s$^{-1}$, meaning that the leptons would be escaping their confinement propagating into the ejecta. This scenario needs to be further explored by conducting an energy-dependent analysis of the morphology for this source, estimating upper limits for the magnetic field amplitude and understanding the degree of turbulence of this system.

\section{Conclusion}
	In this proceeding we have discussed the morphology of the composite SNR MSH 15-5\textit{2} in the VHE band using H.E.S.S. I data (Sect.~\ref{configuration}). We also used a synchrotron X-ray template in the [4 - 7] keV band in order to track the emission in $\gamma$-rays via IC scattering (Sect.~\ref{xrays}). We have found that our best model for adequately fitting the shape of the source, is the model "\textit{v1R+G}" (see Sect.~\ref{v1r+} and Table \ref{table_results}), in which the X-ray emission is linked to the assumption on the magnetic field depending on the projected distance as something similar to $B \propto$ $R^{-\frac{1}{2}}$ and the more extended $\gamma$-ray emission is modelled by an added Gaussian of $\sigma \sim 7'$, bearing 65$\%$ of the total flux.
	
	Furthermore we present some scenarios regarding the behaviour on the leptonic population of the system, based on the characteristic age, size and shape of the SNR MSH 15-5\textit{2} and the pulsar PSR B1509-58 within. We discuss the extent of the interaction of the reverse shock with the medium based on its density gradients (Sect.~\ref{shock}), the spectra for the different components of the source (Sect.~\ref{spectra}) and the possibility of having an energy-dependent morphology for this object when combined to an estimation for the magnetic field in the medium could support the hypothesis of leptons escaping onto the SN ejecta (Sect.~\ref{escape}).
	
	To conclude, MSH 15-5\textit{2} has a quite telling morphology that combined with the spectral information may provide ground for a modelling of the leptonic population within the SNR and the included PWN. A full modelling of the accelerated particles with assumptions on the ambient photon density, the magnetic field behaviour and the interaction of the reverse shock with the ejecta could provide in the near future an improved understanding of this system.

\end{document}